\documentclass{aa}
\usepackage[varg]{txfonts}
\usepackage{psfig}

\begin{document}

\titlerunning{LIV and $\gamma$-ray spectra of blazars}

\title{On the detectability of Lorentz invariance violation through anomalous multi-TeV $\gamma$-ray spectra of blazars}

\author{F. Tavecchio \and G. Bonnoli}

\institute{INAF -- Osservatorio Astronomico di Brera, Via E. Bianchi 46, I--23807 Merate, Italy} 


\abstract
   {Cosmic opacity for very high-energy gamma rays ($E>10$ TeV) due to the interaction with the extragalactic background light can be strongly reduced because of  possible Lorentz-violating terms in the particle dispersion relations expected, e.g., in several versions of quantum gravity theories.}
   {We discuss the possibility to use very high energy observations of blazars to detect anomalies of the cosmic opacity induced by LIV, considering in particular the possibility to use -- besides the bright and close-by BL Lac Mkn 501 -- {\it extreme} BL Lac objects.}
   {We derive the modified expression for the optical depth of $\gamma$ rays considering also the redshift dependence and we apply it to derive the expected high-energy spectrum above 10 TeV of Mkn 501 in high and low state and the extreme BL Lac 1ES 0229+200.}
   {We find that, besides the nearby and well studied BL Lac Mkn 501 -- especially in high state --, suitable targets are {\it extreme} BL Lac objects, characterized by quite hard TeV intrinsic spectra likely extending at the energies relevant to detect LIV features.}
   {}


\maketitle 

\section{Introduction}

Standard Model of particle physics and General Relativity are thought to be low-energy limits of a more fundamental physical theory. Efforts in building such a comprehensive theory often lead to schemes in which the Lorentz invariance is violated at very high energies (e.g., Mattingly 2005, Liberati 2013). Effects related to Lorentz invariance violation (LIV) are expected to be greatly suppressed at low energy by terms of the order $(E/E_{\rm LIV})^n$, where $E$ is the considered energy and $E_{\rm LIV}$ is the relevant energy scale, commonly assumed to be of the order of the Planck energy, $E_{\rm LIV}\approx E_{\rm PL}=\sqrt{\hbar c^5/G}\simeq 1.22\times 10^{19}$ GeV. Although the  effects induced by LIV are  expected to be quite small at energies reachable by most of the current experiments, they can result in observable anomalies in processes characterized by well defined energy thresholds. Indeed, LIV terms modify the standard energy-momentum relation and can induce variations in the kinematics of  scattering and decay processes (e.g., Coleman \& Glashow 1999, Jacobson et al. 2003), both allowing reactions forbidden by the standard physics (e.g. photon decay) and changing energy thresholds, as in the case the $\gamma \gamma \to e^+ e^-$ pair production reaction (e.g. Kifune 1999, 2014; Protheroe \& Meyer 2000). 

The modification of the $\gamma \gamma \to e^+ e^-$ scattering can be effectively probed by  observations of blazars at very high energy. Indeed LIV effects in this reaction become relevant at energies $E\approx (m_{\rm e}^2 c^4 E_{\rm PL}^{n-2})^{1/n}\sim 10$ TeV for $n=3$, the lowest order interesting for deviations in the high-energy regime. Gamma rays of these energies are effectively absorbed through the interaction with the low energy radiation of the extragalactic background light (EBL). Deviations of the scattering kinematic induced by LIV can lead to the reduction of the cosmic opacity, thus allowing high-energy photons ($E>10$ TeV) to evade absorption and reach the Earth. The detection of such opacity anomalies is still difficult, since the performances of current TeV observatories do not allow us to obtain good quality spectra of blazars at energies exceeding 10 TeV. The upcoming Cherenkov Telescope Array (Acharya et al. 2013) and its precursors, such as the proposed ASTRI/CTA mini array (Di Pierro et al. 2013), will greatly improve the  sensitivity above 10 TeV,  providing the ideal instruments to probe and constrain LIV scenarios\footnote{Recently, Kifune (2014) stressed the fact that LIV effects can also influence the formation of showers in the atmosphere through which TeV photons are detected.}. The search for LIV effects through the modification of the cosmic opacity is complementary to the method based on the measure of energy-dependent photon time of flight from cosmic sources (Amelino-Camelia et al. 1998, Ellis \& Mavromatos 2013), recently applied to GRB and blazars and already providing interesting lower limits for the LIV energy scale of photons, $E_{\rm LIV}>9.3\times 10^{19}$ GeV for $n=1$ and $E_{\rm LIV}>1.3\times 10^{11}$ GeV for $n=2$ (Vasileiou et al. 2013). In fact, not all scenarios including LIV predict the same effects and thus different methods can probe LIV in different frameworks. Furthermore, while time-of-flight measurements only test LIV with photons, the method based on the modification of the kinematics of the $\gamma \gamma \to e^+ e^-$ reaction also involves Lorentz violating terms for dispersion relations of the resulting leptons (Kifune 1999).

Recently, Fairbairn et al. (2014) performed a feasibility study concerning the possible detection of spectral LIV effects with  high-energy observations of blazars by CTA, based on the expected opacity for different values of the LIV energy scale for photons. In this Note we extend their treatment to include the redshift-dependence of the EBL and we compare their treatment with the that previously presented by Jacob \& Piran (2008). We further proceed discussing the targets most suitable to be used for this study, proposing that  LIV effects could be effectively constrained through deep observations of the the so-called {\it extreme} BL Lacs (e.g. Costamante et al. 2001, Tavecchio et al. 2011, Bonnoli et al. 2015) located at relatively large redshift ($z\sim 0.1-0.2$).  In \S 2 we review the calculation of the modified optical depth, extending the Fairbairn et al. (2014) treatment. In \S 3 we then discuss the best cases of blazars to be used for probe LIV through the opacity anomaly and finally in \S 4 we conclude.

Throughout the paper, the following cosmological  parameters are assumed: 
$H_0=70$ km s$^{-1}$ Mpc$^{-1}$, $\Omega_{\rm M}=0.3$, $\Omega_{\Lambda}=0.7$.

\section{Gamma-ray absorption with LIV}

Inspired by effective field theories and quantum gravity theories, the modifications that LIV introduces to reaction thresholds at high energy are commonly studied using phenomenological dispersion relations for the involved particles, in which the effects of LIV are expressed by the addition of terms of the form $E^{n+2}/E_{\rm LIV}^n$, where $E$ is the particle energy (e.g., Kifune 1999). We limit the following treatment to the $n=1$ case, for which the modified relation
for photons reads:
\begin{equation}
E_{\gamma}^2=p_{\gamma}^2c^2 -  \frac{E_{\gamma}^3}{E_{\rm LIV}}
\label{dispersion}
\end{equation}
\noindent
where $p_{\gamma}$ is the photon momentum.
The term $m^2_{\gamma} \equiv- E_{\gamma}^3/E_{\rm LIV}$ acts as an effective mass term for photons and induces modifications of the threshold of the $\gamma \gamma \to e^+ e^-$ scattering for energies in which the term  becomes comparable to the threshold energy $\approx m_{\rm e}c^2$. In principle, the LIV terms can assume different values for different particle species\footnote{In fact we neglect possible LIV terms for electrons.} and can be both positive or negative. The sign assumed above is that for which an interesting anomaly  (i.e. decreasing opacity for increasing energy) is displayed for every value of $E_{\rm LIV}$. 
The $E_{\rm LIV}$ term in the denominator -- related to the energy scale of LIV effects -- is generally assumed to be of the order of the Planck energy. 

The calculation of the modified optical depth including LIV effects is quite delicate, since in the LIV framework, even basic standard assumptions (e.g. energy-momentum conservation) could be not valid. In the literature, expressions for the optical depth, derived with two alternative assumptions, are given by Jacob \& Piran (2008) and  Fairbairn et al. (2014). In both works the modified dispersion relation given in Eq.\ref{dispersion} is assumed and the same modified expression for the minimum energy of the soft target photons allowing the pair production reaction is found (see also Kifune 1999):
\begin{equation}
\epsilon_{\rm  min}= \frac{m_{\rm e}^2c^4}{E_{\gamma}} + \frac{E^2_{\gamma}}{4 E_{\rm LIV}} 
\label{eq:epsmin}
\end{equation}
where the last term is introduced by LIV effects. This expression can be obtained solely on the minimal assumption that the standard energy-momentum conservation still holds in a LIV framework (note, however, that this could be not true in some LIV schemes, most notably in the so-called double special relativity, Amelino-Camelia et al. 2005).

\begin{figure}
\psfig{file=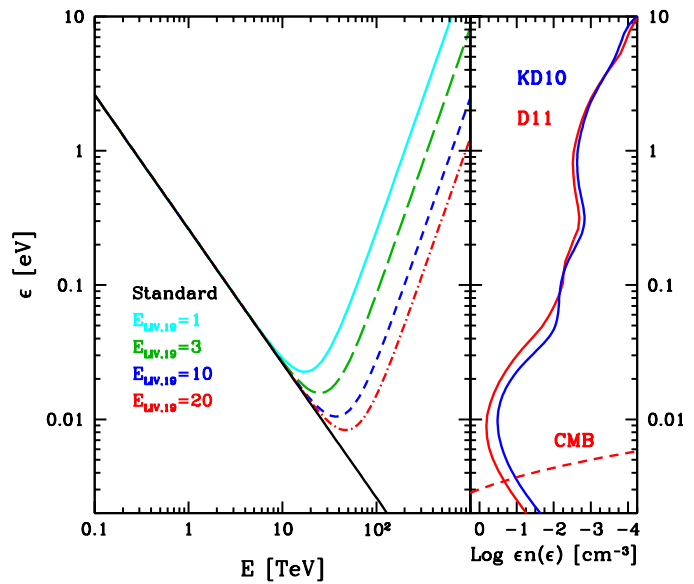,height=9cm,width=9.cm}
\vspace{-.7 truecm}
\caption{{\it Left}: photon target energy at threshold $\epsilon_{\rm min}$ for the pair-production reaction as a function of the $\gamma$-ray energy $E_{\gamma}$. The black solid line shows the standard case. The other lines report the modified threshold resulting from the LIV modified kinematics, for different values of the parameter $E_{\rm LIV}$ (in units of $10^{19}$ GeV). {\it Right}: EBL (Dominguez et al. 2011, solid red; Kneiske \& Dole 2010, solid blue) and CMB (dashed red) local density $\epsilon n(\epsilon)$ (horizontal axis) as a function of the energy $\epsilon$ (vertical axis), to be compared with the energy threshold in the left panel.
}
\label{epsilon}
\end{figure}

The resulting energy $\epsilon_{\rm min}$ of target photons at threshold for the reaction with gamma rays with energy $E_{\gamma}$ is shown in Fig. \ref{epsilon}. The black solid line reports the standard value $\epsilon _{\rm min}=m_{\rm e}^2c^4/E_{\gamma}\simeq 0.26/E_{\gamma, TeV}$ eV. The other lines show the modified threshold energy for different values of $E_{\rm LIV}$, normalized to the expected characteristic energy for LIV, $E_{\rm LIV}=10^{19}$ GeV. The common feature of the curves including LIV effects is the existence of a minimum for $\epsilon_{\rm min}$, corresponding to energies of the incoming $\gamma$ ray around $E_{\rm c}=30-50$ TeV. The existence of the minimum implies a progressive reduction of the resulting optical depth above $E_{\rm c}$. The curves corresponding to increasing value of $E_{\rm LIV}$ (from $E_{\rm LIV}=10^{19}$ GeV to $2\times 10^{20}$ GeV) track the progressive shift of $E_{\rm c}$ to higher energies.  

The treatments of Jacob \& Piran (2008) and  Fairbairn et al. (2014) differs from the assumptions made to further derive the expression for the optical depth $\tau_{\gamma \gamma}$ (see the Appendix for more details.). Jacob \& Piran (2008) assume the standard formula (e.g. Dweck \& Krennrich 2012) and that the functional form of the cross-section -- that in the standard case can be expressed as a function of the ratio $\epsilon/\epsilon _{\rm min}$ -- still holds at energies where LIV effects begin to be important. It is worth to note that their procedure avoid any change of reference frame, since any quantity is evaluated in the observer frame. The optical depth can thus be calculated with the standard expression.

Fairbairn et al. (2014) instead used a modified expression for the square of the center-of-mass energy $s$ including the effective mass for the photon $m^2_{\gamma}$,  $s=m^2_{\gamma}+2\epsilon E_{\gamma}(1-\cos \theta)$, in which $\theta$ is the angle between the two photon directions, and assumed that $s$ continues to be a  good invariant (see Jacob et al. 2010 for a justification of this assumption). 
They calculated the expected modification of the cosmic opacity for gamma rays for close-by sources, for which the redshift and the EBL evolution can be safely neglected.
In the following calculation of the optical depth we use a redshift-dependent EBL density, which allows us to consider sources at arbitrary distance.

The standard relation for optical depth at the energy $E_{\gamma}$ and for a source at redshift $z_{\rm s}$ (e.g. Dwek \& Krennrich 2013) is modified as (Fairbairn et al. 2014):
\begin{multline}
\tau_{\gamma\gamma}(E_{\gamma},z_{\rm s})=\frac{c}{8E_{\gamma}^2}\int_0^{z_{\rm s}} \frac{dz}{H(z)(1+z)^3}\, \int_{\epsilon_{\rm min}(z)}^{\infty} \frac{n(\epsilon,z)}{\epsilon^2}d\epsilon \\
 \int_{s_{\rm min}(z)}^{s_{\rm max}(z)} [s-m^2_{\gamma}(z)] \, \sigma _{\gamma\gamma}(s)\, ds
\label{eq:tau}
\end{multline}
\noindent
where $H(z)=H_{0}\, [\Omega _{\Lambda}+\Omega_{M}(1+z)^3]^{1/2}$, $\epsilon$ is the target photon energy, $n(\epsilon,z)$ is the redshift-dependent differential EBL photon number density, $\sigma_{\gamma\gamma}(s)$ is the total pair production cross section as a function of the modified square of the center of mass energy $s=m^2_{\gamma}+2\epsilon E_{\gamma}(1-\cos \theta)$. The limits of the last integral reads:
\begin{equation}
s_{\rm  min}=4m_e^2c^4 
\label{smin}
\end{equation}
\begin{equation}
s_{\rm  max}=  4\epsilon E_{\gamma}(1+z) - \frac{E^3_{\gamma}(1+z)^3}{E_{\rm LIV}}.
\label{smax}
\end{equation}


Standard relations are clearly recovered for $E_{\rm LIV}\to \infty$. In Eq. \ref{smax} we have neglected the energy-dependent speed of light $\beta_{\gamma}(E_{\gamma})$, since for the energies relevant here $\beta_{\gamma}(E_{\gamma})\simeq 1$. The $(1+z)$ terms in Eq. 4-5 which take into account the progressive redshift of the $\gamma$ rays while they propagate from the source to the Earth.

For the EBL density $n(\epsilon,z)$ we use two models, namely  the state-of-the art model by Dominguez et al. (2011) (D11 hereafter) and the model by Kneiske \& Dole (2010) (KD10). The latter model predicts a somewhat lower level of IR radiation, determining a smaller optical depth at energies above few TeV. The local photon densities predicted by the two models are reported in the right panel of Fig. 1

\begin{figure}
\hspace{-.9 truecm}
\psfig{file=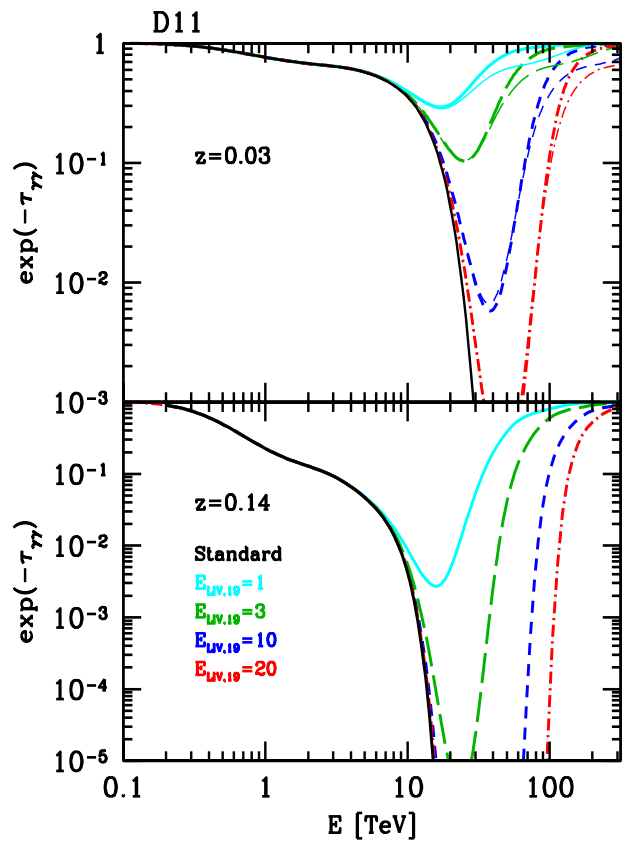,height=12.cm,width=11.cm}
\caption{Absorption coefficient $e^{-\tau_{\gamma\gamma}}$ as a function of  energy for $\gamma$ rays propagating from a source at $z=0.03$ (upper panel) and $z=0.14$ (lower panel) using the EBL model of Dominguez et al. (2011). The black solid line refers to the standard case, the other lines show the modified coefficient for different value of $E_{\rm LIV}$ (with the color code reported in the caption). Thick lines have been obtained using the treatment of Fairbairn et al. (2014). For comparison, thin lines report the results of the calculations based the assumptions of Jacob \& Piran (2008).}
\label{tau}
\end{figure}

\begin{figure}
\hspace{-.9 truecm}
\psfig{file=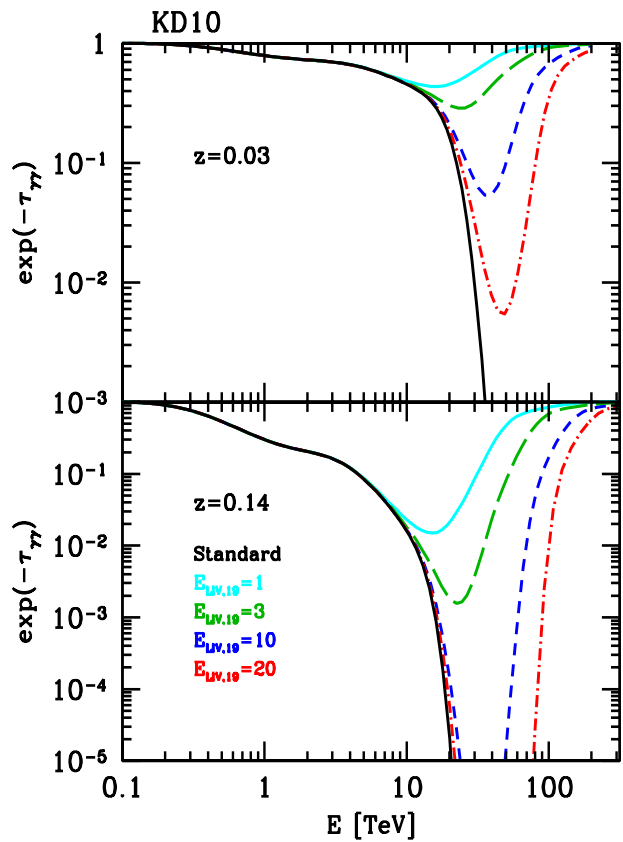,height=12.cm,width=11.cm}
\caption{As Fig. \ref{tau} but using the EBL model of Kneiske \& Dole (2010).}
\label{tau2}
\end{figure}

In Figs. \ref{tau}-\ref{tau2} we report (thick lines) the absorption coefficient $\exp[-\tau_{\gamma \gamma}(E_{\gamma})]$ (calculated with Eq. \ref{eq:tau}) for two values of the redshift, $z=0.03$ (upper panel) and 0.14 (lower panel) and the values of $E_{\rm LIV}$  considered above and the two different EBL models. The drastic reduction of the opacity above few tens of TeV induced by the LIV effect is clearly visible. 
In the top panel of Fig. \ref{tau} we also report the curves (thin lines) corresponding to optical depths evaluated with the method of Jacob \& Piran (2008). Clearly, the results of the two methods differ only at the highest energies and the ratio between the two absorption coefficients is always less then a factor of two. For simplicity, in the following we only report the results obtained with the Fairbairn et al. (2014) treatment. See the Appendix for more details.

\section{Application to blazar spectra}

An ideal source to test possible modification of the gamma-ray opacity induced by LIV should be a bright emitter above 20-30 TeV, at which the LIV effects become fully appreciable. Unfortunately, this requirement is in conflict with the typical characteristics of the VHE emitting blazars, which commonly display (intrinsic)  spectra softening with energy, as a result of the decreasing inverse Compton scattering and particle acceleration efficiencies and, possibly, internal opacity. However, given the extreme variability characterizing blazars, it is not unlikely that some sources can display hard and bright TeV emission particularly suitable for the present analysis. It is also becoming clear that a class of peculiar BL Lacs, known as extreme BL Lacs (Costamante 2001, Tavecchio et al. 2011, Bonnoli et al. 2015),  are characterized by a very hard and stable TeV continuum, possibly extending above 10 TeV. The relatively high redshift of these sources ($z>0.1$) implies a relatively large absorption, which is however more than compensated by the expected intrinsic flux above 20-30 TeV, where LIV effects become important. 

With these motivations, in the following we investigate the prospects to probe LIV spectral effects using observations of high-states of Mkn 501, a classical TeV BL Lac objects, and of the prototype extreme BL Lac 1ES 0229+200.

\begin{figure}
\hspace{-0.45 truecm}
\psfig{file=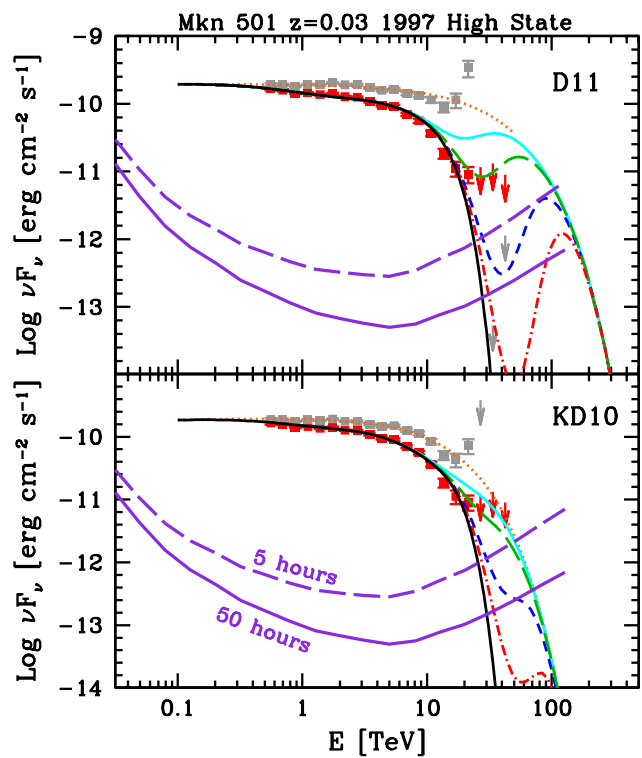,height=10.5cm,width=10.5cm}
\caption{Observed $\gamma$-ray spectrum of Mkn 501 during the 1997 active state recorded HEGRA (red symbols). The black solid and the dotted orange lines report the observed and the intrinsic spectrum assuming standard absorption with the D11 (upper panel) and KD10 (lower panel) EBL models. Grey points show the observed data points corrected for absorption. The other lines show the predicted spectrum with LIV effects (line styles and colors as in Fig. \ref{tau}). The  violet lines report the $5\sigma$ sensitivity curves for  CTA (5 hours, dashed, and 50 hours exposure, solid). 
}
\label{501h}
\end{figure}

\subsection{Mkn 501}

Early studies (e.g., Kifune 1999, Protheroe \& Meyer 2000) focused on the close-by and luminous BL Lac objects Mkn 501. The spectrum of this source during quiescent states has been also considered by Fairbairn et al. (2014) to investigate the CTA potentialities to detected LIV. A problem with this and similar sources (e.g., Mkn 421) is the typical steep spectrum, which implies that the value of the intrinsic flux above 20-30 TeV is expected to be quite low. Moreover, current observations, limited to $E\lesssim20$ TeV, do not ensure that the emission continues at the required energies without breaks. A steepening or a cut-off of the emitted spectrum could indeed hamper or strongly limit the application of the method. 

Quite interestingly, Mkn 501 occasionally shows active states in which the spectrum becomes remarkably hard (photon index $\Gamma_{\rm VHE}\sim 2$) and extends  at least up to $\sim20$ TeV. The most studied high state occured in 1997 and for such state there is a superb quality spectrum recorded by HEGRA (Aharonian et al. 1999). Other two such states have been observed in 2009 and 2011 and VHE spectra have been obtained by VERITAS (Abdo et al. 2011) and ARGO (Bartoli et al. 2012). As shown by Neronov et al. (2012) the hard TeV spectrum in 2009 was also accompanied by an exceptionally hard spectrum above 10 GeV detected by LAT ($\Gamma _{\rm LAT}\sim 1$). These active phases lasted for about several weeks. Clearly, the spectral hardness and the high flux during these flaring states would be ideal to study LIV effects. In the following we use the HEGRA 1997 spectrum.


In Fig. \ref{501h} we show our predictions for the observed spectrum at high energy during this activity state  for the two EBL models. The choice of the spectral shape (a power-law with exponential cut-off) is constrained by the requirement to reproduce the {\it observed} data points (red symbols) assuming no LIV. In all cases, in presence of LIV effects the spectrum is predicted to show a quite narrow upturn, where the observed spectrum would recover to the intrinsic one.  

 Note that, clearly, the predicted spectra for the D11 EBL model and the cases with  $E_{\rm LIV}=10^{19}$ GeV and $3\times 10^{19}$ GeV are inconsistent with the observed data. Therefore, these data for Mkn 501 already suggest $E_{\rm LIV}\gtrsim 3\times 10^{19}$ GeV.

 For comparison, in Fig. \ref{501h} we show the expected differential $5\sigma$ sensitivity curves for CTA with an exposure time of 50 h (from Bernl{\"o}hr et al. 2013, model ``SAM" in Fig. 14) and 5 h (derived from Fig. 8 in  Bernl{\"o}hr et al. 2013).
With the longest assumed exposure, in the D11 case, CTA should be able to easily reveal the LIV upturn even for the largest assumed value for the photon LIV parameter, $E_{\rm LIV}=2\times 10^{20}$ GeV. For the case of the KD10 EBL model, the required intrinsic spectrum has a cut-off at an energy lower than that found for the D11 case, since the absorption around 10 TeV is quite lower. In turn, a lower energy cut-off implies that the LIV upturn is much less pronounced , making difficult to probe values of $E_{\rm LIV}$ larger than $\approx10^{20}$ GeV

\begin{figure}
\hspace{-2.7 truecm}
\psfig{file=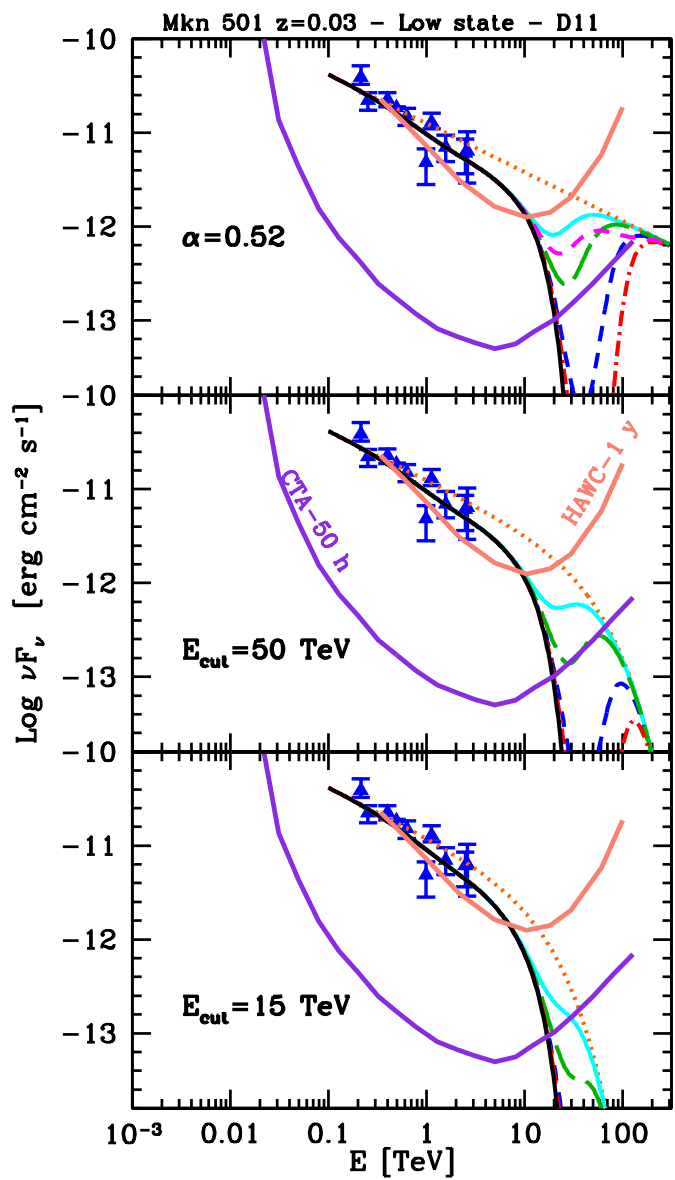,height=15.5cm,width=15.5cm}
\caption{Observed $\gamma$-ray spectrum of Mkn 501 during a quiescent phase (Acciari et al. 2011). The three panels report the received $\gamma$-ray spectrum for different assumed intrinsic spectra (from top to bottom: a simple power law with energy index $\alpha=0.52$; a power law with energy index $\alpha=0.52$ and an exponential cut-off with $e-$folding energy $E_{\rm cut}=50$ TeV; a cut-offed power  law with  $\alpha=0.52$ and $E_{\rm cut}=15$ TeV) and values of $E_{\rm LIV}$ (color code as in Fig. 2). The HAWC and CTA sensitivity curves (as in Fig. 3) are displayed for an exposure time of 1 year and 50 hours, respectively.
}
\label{501l}
\end{figure}

\begin{figure}
\hspace{-2.7 truecm}
\psfig{file=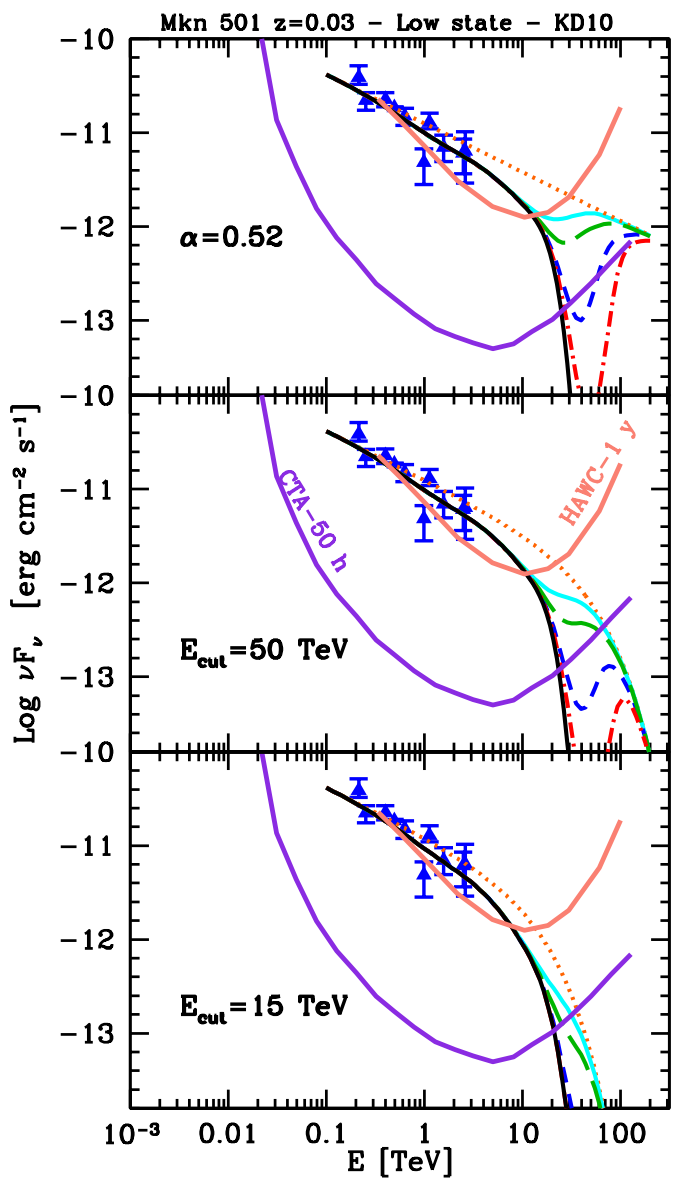,height=15.5cm,width=15.5cm}
\caption{As Fig. \ref{501l} but with the KD10 EBL model.
}
\label{501lKD}
\end{figure}

For comparison, Figs. \ref{501l}-\ref{501lKD} report the case corresponding to a low state -- the same assumed by Fairbairn et al. (2014). We note in passing that they incorrectly assumed that the intrinsic spectral slope is that traced by the observed spectrum. On the contrary,  as noted above, the optical depth can not be neglected and indeed the intrinsic spectrum is harder than the observed one, with $\Delta \alpha\simeq 0.2$. This is valid for both considered EBL models, since the observed datapoints cover an energy range for which the two models basically provide the same opacity. Once extrapolated at high-energy, the flux is thus larger than what assumed in Fairbairn et al. (2014): their results should thus be considered somewhat pessimistic. This is particularly important for the KD10 model, which provides a lower opacity and thus a large flux at energies where LIV effects are important (note the difference with the previous case of the high state, for which the {\it observed} spectrum extends up to energies where the KD10 opacity is lower and thus the predicted flux for LIV effects is lower than the D11 case).
 
With 50 hours, CTA can still easily detect the spectral upturn for all parameters assumed here in the case in which the intrinsic emission follows an unbroken power-law. If the spectrum exponentially drops with $E_{\rm cut}>50$ TeV the upturn can still be detected in the most favorable cases. For smaller values of $E_{\rm cut}$ the detection of the anomalous transparency is challenging. For comparison we also report the differential sensitivity curve 1 year exposure with HAWC (Abeysekara et al. 2013). In fact, quiescent states are likely to last for long time and thus in this case we can assume relatively long exposure times, suitable also for HAWC and other instruments (see Discussion).

\subsection{Extreme Highly peaked BL Lacs: 1ES 0229-200}

As already stressed, and as clearly shown by the last example, the effective investigation of LIV spectral anomalies benefits from hard spectra and large maximum energies. In view of these requirements, extreme BL Lacs (EHBL), characterized by quite hard TeV spectra extending at least up to 10 TeV, are expected to be good targets. The peculiar emission properties of these sources are still not clearly understood. EHBL are characterized by extremely low radio luminosity together with luminous and hard X-ray emission, often locating the peak of the synchrotron emission above 10 keV (e.g. Bonnoli et al. 2015). The hardness of the de-absorbed VHE spectrum is challenging for the standard one-zone leptonic model, in which the inverse Compton scattering of multi-TeV electrons -- which in this framework is responsible for the $\gamma$-ray emission -- becomes quite inefficient (Katarzy{\'n}ski et al. 2006, Tavecchio et al. 2009, 2011, Kaufmann et al. 2011). Another peculiarity is related to the absent or very weak variability displayed by the VHE emission (e.g., Aliu et al. 2014), at odds with the typical extreme behavior of the bulk of the BL Lac population. Possible alternative explanations include hadronic emission (Cerutti et al. 2015, Murase et al. 2012), internal absorption (Zacharopoulou et al. 2011) or quasi-Maxwellian energy distribution of the emitting leptons (Lefa et al. 2011). 

\begin{figure}
\hspace{-2.5 truecm}
\psfig{file=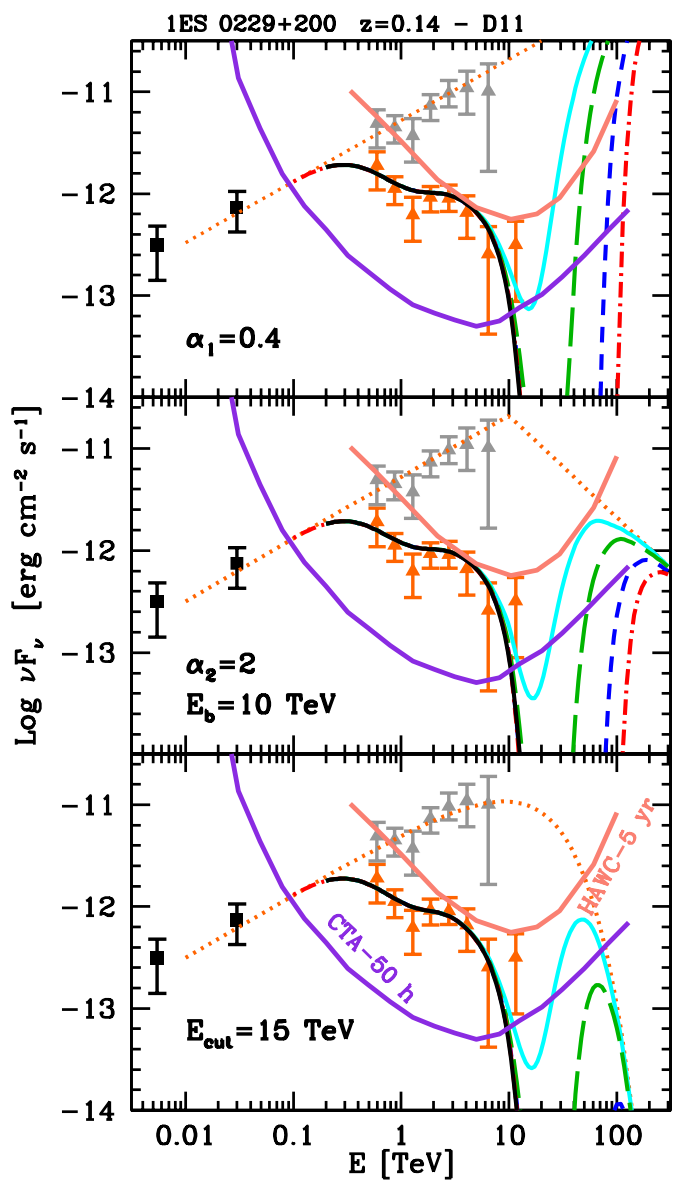,height=15.5cm,width=15.5cm}
\caption{Observed high-energy SED of 1ES 0229+200. Black symbols are LAT measurements from the 3FGL catalogue. Orange triangles mark the HESS spectrum (Aharonian et al. 2007), while the gray triangles are obtained after correction for the absorption with EBL. The dotted line indicates the assumed intrinsic spectrum. The three panels report the received $\gamma$-ray spectrum for different assumed intrinsic spectra (from top to bottom: a simple power law with energy index $\alpha=0.4$; a broken power law with high-energy index $\alpha_2=2$ and break energy $E_{\rm b}=10$ TeV; an exponential cut-off with $e-$folding energy $E_{\rm cut}=15$ TeV) and values of  $E_{\rm LIV}$ (color code as in Fig. 2). The HAWC and CTA sensitivity curves (as in Fig. 3) are displayed for an exposure time of 5 year and 50 hours, respectively.
}
\label{0229}
\end{figure}

\begin{figure}
\hspace{-2.5 truecm}
\psfig{file=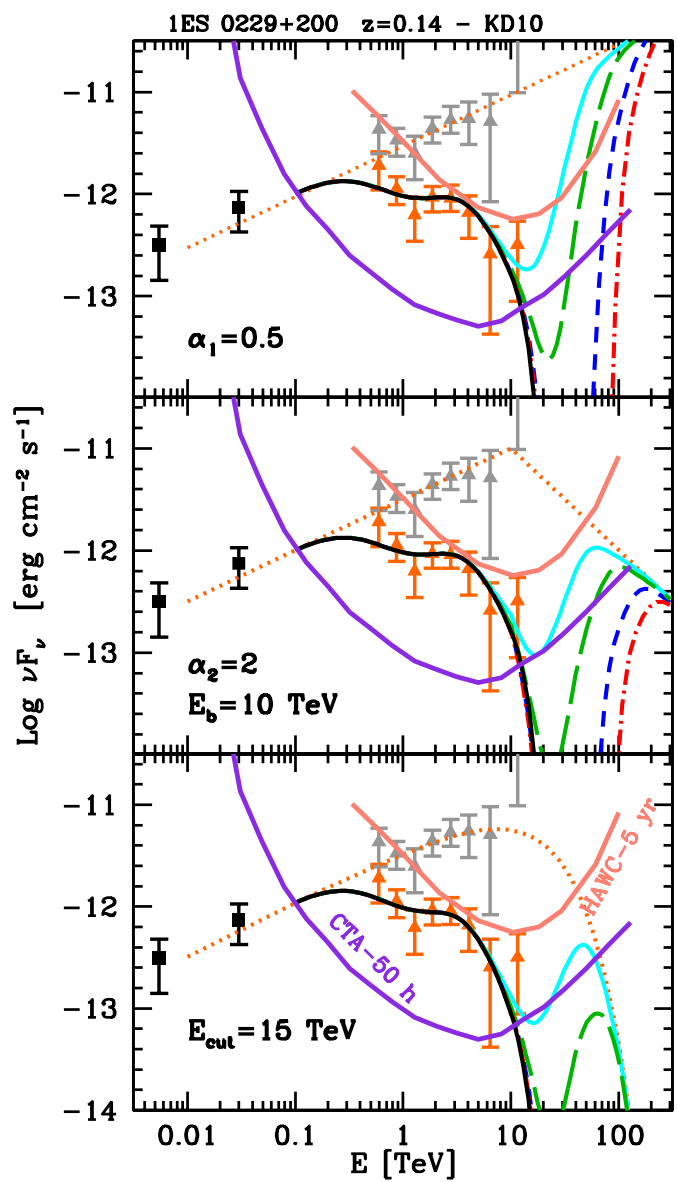,height=15.5cm,width=15.5cm}
\caption{As Fig. \ref{0229} but with the KD10 EBL model and $\alpha=0.5$.
}
\label{0229KD}
\end{figure}

As a benchmark case we consider 1ES 0229-200 ($z=0.14$), the prototype of this class of sources (Tavecchio et al. 2009, Bonnoli et al. 2015). The relatively large redshift of 1ES 0229-200 implies an important absorption of the VHE spectrum. The observed spectrum is indeed soft ($\Gamma\simeq 2.5$) but, correcting for (standard) absorption, results in very hard continuum, apparently unbroken up to 10 TeV (Aharonian et al. 2007). In Figs. \ref{0229}-\ref{0229KD}, together with the observed data points and the those with the correction for absorption with the two EBL models, we present the prediction for the high-energy spectrum based on three possible intrinsic spectral shapes compatible with the observed data, namely a power law with $\alpha=0.4$ (with D11) or $\alpha=0.5$ (with KD10), a broken power law with $\alpha_1=\alpha$, $\alpha_2=2$ and $E_{\rm b}=10$ TeV (the minimum compatible with the data) and a power law with exponential cut-off at $E_{\rm cut}=15$ TeV. As for the case of Mkn 501, for the last two models we assume the lowest value of break and cut off energy compatible with the data, thus providing a conservative estimate.

As above, we compare the predicted fluxes with the sensitivity curves for HAWC and CTA. We remark that the $\gamma$-ray spectrum of 1ES 0229+200 appears to be rather stable, showing only marginal variations on timescales of few weeks (Aliu et al. 2014). In this respect it is an ideal source for prolonged exposures, since the signal can be accumulated over long periods without problems related to important spectral changes. The exposure time for CTA could thus be even larger than those assumed here (50 h).  Since the flux limit at the highest energies -- at which the cosmic-ray background is almost negligible -- is expected to scale linearly with time (e.g., Bernl{\"o}hr et al. 2013), even a doubling of the exposure can lead to an important improvement of the constraints.

However, even with the expected improvements, with the LIV parameters in the range investigated in the present work, CTA is expected to reveal excess flux at high energy only for both the power-law or the broken power law case. An exponential cut-off at $E_{\rm cut}=15$ TeV would imply a large suppression of the flux in the relevant band and only for the smaller LIV parameters (possibly ruled out by the 1997 HEGRA spectrum) the upturn could be detected. Note that the  predictions made with the two EBL models are quite similar, being only slightly more pessimistic with KD10.

\section{Discussion}

We have revisited the possibility to detect  anomalies induced by LIV in TeV spectra of blazars. 
We model the anomalous absorption extending the treatment used by Fairbairn et al. (2014) to sources for which redshift is not negligible. We have also compared the resulting absorption coefficient with that obtained with the alternative approach of Jacob \& Piran (2008). Fairbairn et al. (2014) assumed that the square of the center-of-mass energy $s$ -- modified to include the effective mass of the high-energy photon induced by LIV -- is still an invariant quantity even in presence of LIV. On the contrary, Jacob \& Piran (2008) did not make any assumption on the behavior of $s$, but assumed that the functional from of the pair production cross-section as a function of the ratio $\epsilon/\epsilon_{\rm min}$ holds also in the LIV regime. In spite of the two different assumptions, the resulting absorption coefficients only differs by a small factor in the considered energy range. We remark that a limit of both approaches is that they consider only the modification of the scattering kinematic caused by the modified dispersion relations but they do not adopt any real dynamical scenario  considering the LIV effects on the cross section (e.g. Colladay \& Kosteleck{\'y} 2001, Rubtsov et al. 2012) and just use some educated guesses to extrapolate the cross section in the LIV regime.

While previous studies focus on nearby ``classical " TeV BL Lacs -- whose prototype is Mkn 501 -- we have  emphasized  the possible role of extreme BL Lacs, whose intrinsic hard spectra seems to be ideal for such studies. We further remark that these sources are quite interesting VHE target by themselves for several other reasons: in particular the hard spectrum makes them ideal to probe the EBL deep in the far IR band. Moreover, observations at 20-30 TeV could definitely prove or rule out the intriguing hypothesis  that the peculiar TeV emission could be the result of cosmic ray beamed by the jet toward the Earth (Essey et al. 2011, Murase et al. 2012)  or to prove the existence of axion-like particles mixing with photons (e.g. De Angelis et al. 2011). Observations of these sources can thus address a large spectrum of physical topics.

We have shown that CTA could be effectively used to put strong constraints to LIV for energy scales $E_{\rm LIV}=10^{19}-10^{21}$ GeV. The existing HEGRA data taken during the major outburst of 1997 already seems to exclude  $E_{\rm LIV}<2-3\times 10^{19}$ GeV (see also Biteau \& Williams 2015). Some of the lowest values of $E_{\rm LIV}$, resulting in quite high fluxes at 20-30 TeV, could be also potentially already ruled out by available data from HESS or MILAGRO. We have shown the results for two different EBL models, with that by Kneiske \& Dole (2010) predicting a quite low opacity above 10 TeV.

We remark here that the comparison made in this work between the predicted spectrum and the expected differential sensitivity curves 
can be done more precisely by means of dedicated simulations.
 
Simulations along these lines for the ASTRI mini array (Di Pierro et al. 2013, Vercellone et al. 2013) are already in progress.
We also note that a number of factors could improve the sensitivity at the highest energies, where possible LIV effects can appear. An interesting point to note is that prolongued observations tend to  favor  the highest energies. In fact, while the flux limit increases as $\sqrt{t}$ for low energies, at high energy, where the background is strongly reduced ($E\gtrsim 10$ TeV), the sensitivity is expected to incerase linearly with $t$. Another parameter likely impacting on the sensitivity at the highest energies is the zenith angle of the observation: high ZA, translating in large effective areas, would be particularly favorable for LIV studies.

We would like also to highlight that, besides CTA and HAWC, other instruments could provide interesting results in the search for LIV effects in the gamma-ray spectra of cosmic sources. In particular, HiSCORE (Tluczykont et al. 2014) will extend with good sensitivity (but with quite prolonged observations) the energy range above 100 TeV. Even more promising appears LAAHSO, expected to provide an excellent covering above $E\gtrsim 10$ TeV, reaching (integral) fluxes as low as few $10^{-14}$ erg cm$^{-2}$ s$^{-1}$ around 50-100 TeV for 1 year exposure (Cui et al. 2014). 
We would like to note that, given that the LIV spectral signatures are quite narrow, this study can take profit of instruments with good spectral resolution (like CTA and HiSCORE).

Finally, we remark that the approach based on detection of spectral anomalies is complementary to the method based on energy-dependent delays of photons (Amelino-Camelia et al. 1998), whose use
 can be partly hampered by the possible energy-dependent variability of the intrinsic emission induced by acceleration/cooling processes acting on the emitting particles in the jet (e.g., Chiaberge \& Ghisellini 1999, Bednarek \& Wagner 2008). 

\begin{acknowledgements}
We thank A. Giuliani for discussions. We thank the referee for an insightful review that  helped us to greatly improve the paper. This work has been partly founded by a PRIN-INAF 2014 grant. The authors acknowledge joint financial support by the CaRiPLo Foundation and the Regional Government of Lombardia to the project ID 2014-1980 "Science and technology at the frontiers of gamma-ray astronomy with imaging atmospheric Cherenkov Telescopes". 
\end{acknowledgements}

\appendix

\section{Comparison between the Fairbairn et al. and Jacob \& Piran treatments}

In the standard framework, the optical depth for $\gamma$ rays of energy $E$ propagating from a source at distance $d_s$  which interact with the soft EBL photons through the $\gamma\gamma\to e^{+}e^{-}$ reaction can be written as:
\begin{equation}
\tau(E)=\int_{0}^{d_{\rm s}} \int _{\epsilon_{\rm min}} n(\epsilon) \int _{-1}^{1} \frac{(1-\mu)}{2} \, \sigma_{\gamma\gamma}(\beta) \,d\mu\, d\epsilon \, dl
\label{tauapp}
\end{equation}
in which the second integral is performed over the soft photon energies starting from  $\epsilon_{\rm min}$ -- dictated by the energy threshold -- and the third integral is performed over all the incident angles $\theta$ ($\mu=\cos \theta$). 
The total cross section has the expression:
\begin{equation} 
\sigma_{\gamma \gamma}(\beta) =  \frac{\pi r_e^2}{2}\left(1-\beta^2 \right) \left[ 2 \beta \left( \beta^2 -2 \right) + \left( 3 - \beta^4 \right) \, {\rm ln} \left( \frac{1+\beta}{1-\beta} \right) \right],
\label{eq.sez.urto}
\end{equation}
which depends upon $E$, $\epsilon$ and $\mu$ only through the dimensionless parameter:
\begin{equation} 
\beta(s) \equiv \left[ 1 - \frac{4 \, m_e^2 \, c^4}{s} \right]^{1/2}, 
\label{beta}
\end{equation}
where $s$ is the invariant square of the center-of-momentum energy which, using lab quantities is $s=2\epsilon E (1-\mu)$.

The introduction of a LIV framework leads to change Eq. \ref{tauapp} considering the modification of the threshold energy $\epsilon_{\rm min}$ and the parameter $\beta$.

Quite generally, the modified value of the threshold can be derived resorting to conservation of energy and momentum in the observer frame and leads to Eq. \ref{eq:epsmin}. Instead, the modifications to the cross section are not so straightforward to evaluate (see e.g. Colladay \& Kosteleck{\'y} 2001, Rubtsov et al. 2012). The treatments of Jacob \& Piran (2008) and Fairbairn et al. (2014) differ from the basic guess on this point.

Jacob \& Piran (2008) used the fact that, in the standard framework in which $\epsilon _{\rm min}=2m_e^2c^4/E(1-\mu)$, $\beta$ can be re-written as:
\begin{equation} 
\beta(\epsilon/\epsilon_{\rm min}) =  \left[ 1 - \frac{\epsilon_{\rm min}}{\epsilon} \right]^{1/2}.
\label{betajp}
\end{equation}
Then, they made the basic assumption that {\it the same expression is still valid in the LIV framework}, assuming that $\epsilon_{\rm min}$ is given by the  modified LIV expression, Eq. \ref{eq:epsmin}. Note that in this way the cross section can be evaluated without any change of reference frame, using only observer measured quantities.

The alternative approach adopted by Fairbairn et al. (2014) is based instead on the assumption that {\it a modified expression of the  center-of-mass energy squared $s$ is still a good invariant quantity even in the LIV framework}. In particular they define $s=m_{\gamma}^2+2E\epsilon(1-\beta_{\gamma} \mu)$, formally treating $m^2_{\gamma}$ as an effective mass of the high-energy photon and in which $\beta_{\gamma}\lesssim1$ is the high-energy photon velocity. To apply their scheme, Fairbairn et al. (2014) then express the integral over the incident angle in Eq. \ref{tauapp} as an integral over $s$. Making the approximation $\beta_{\gamma}=1$, we have $(1-\mu)=(s-m_{\gamma}^2)/2E\epsilon$ and $d\mu=-ds/2\epsilon E$ and thus the new expression is:
\begin{multline}
\tau_{\rm F}(E)=\int_{0}^{d_{\rm s}} \int _{\epsilon_{\rm min}} \frac{n(\epsilon)}{8E^2\epsilon^2}\int _{s_{\rm min}}^{s_{\rm max}} (s-m^2_{\rm \gamma}) \, \sigma_{\gamma\gamma}(\beta_{\rm F}) \,ds \, d\epsilon \, dl = \\
\int_{0}^{d_{\rm s}} \int _{\epsilon_{\rm min}} \frac{n(\epsilon)}{8E^2\epsilon^2}\,  I_{\rm F}(\epsilon) \, d\epsilon \, dl
\label{tauappf}
\end{multline}
where $s_{\rm min}=4m_e^2c^4$, $s_{\rm max}=4E\epsilon + m_{\gamma}^2$ (which correspond to the previous limit $\mu=-1$) and $\beta_{\rm F}$ is evaluated according to Eq. \ref{beta}.

\begin{figure}
\hspace{-0.5 truecm}
\psfig{file=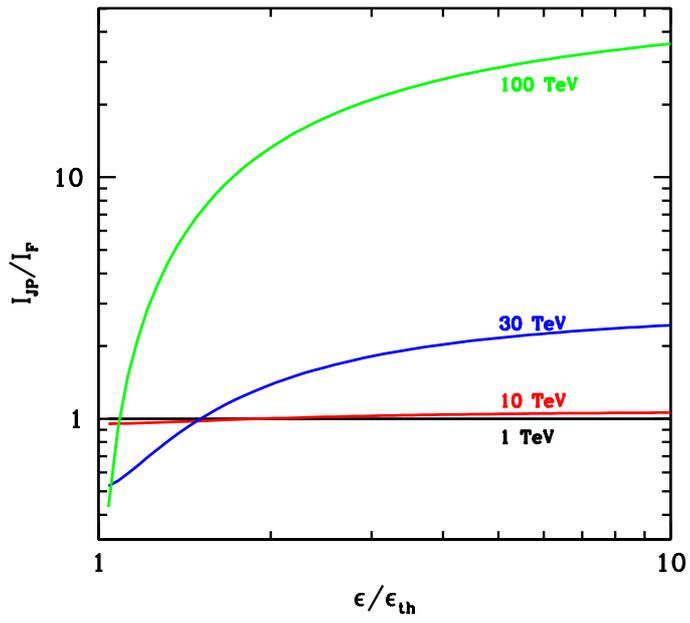,height=10.5cm,width=10.5cm}
\vspace{-1.1 cm}
\caption{Ratio of the two integrals $I_{\rm JP}({\epsilon})$ and $I_{\rm F}({\epsilon})$ as a function of $\epsilon$ for $E_{\rm LIV}=10^{19}$ GeV and for different values of the gamma ray energy, $E=1$ TeV (black), 10 TeV (red), 30 TeV (blue) and 100 TeV (green).  
}
\label{fig:appendix}
\end{figure}

It is possible to check that the two different approaches results in different values of the optical depth. In particular, the Fairbairn et al. treatment provides smaller optical depth for energies above the onset of LIV effects. This can be seen recasting Eq.\ref{tauapp} used by JP in the same form of Eq. \ref{tauappf} making the formal change of variable $\mu \to \tilde{s}=m^2_{\gamma}+2E\epsilon(1-\mu)$ (we remark that we are not attributing any physical meaning to this quantity) but maintaining Eq. \ref{betajp} for the argument of $\sigma_{\gamma\gamma}$. Therefore we obtain:
\begin{multline}
\tau_{\rm JP}(E)=\int_{0}^{d_{\rm s}} \int _{\epsilon_{\rm min}} \frac{n(\epsilon)}{8E^2\epsilon^2}\int _{\tilde{s}_{\rm min}}^{\tilde{s}_{\rm max}} (\tilde{s}-m^2_{\rm \gamma}) \, \sigma_{\gamma\gamma}(\beta_{\rm JP}) \,d\tilde{s} \, d\epsilon \, dl = \\
\int_{0}^{d_{\rm s}} \int _{\epsilon_{\rm min}} \frac{n(\epsilon)}{8E^2\epsilon^2}\,  I_{\rm JP}(\epsilon) \, d\epsilon \, dl
\label{tauappjp}
\end{multline}
where, using Eq. \ref{eq:epsmin} for $\epsilon_{\rm min}$ to expand Eq. \ref{betajp},
\begin{equation} 
\beta_{\rm JP}(\tilde{s}) =  \left[ 1 - \frac{\epsilon_{\rm min}}{\epsilon} \right]^{1/2} = \left[ 1 - \frac{4m_e^2c^4-m_{\gamma}^2}{\tilde{s}-m^2_{\gamma}} \right]^{1/2}.
\end{equation}
Clearly $\beta_{\rm JP}\neq \beta_{\rm F}$ (Eq. \ref{beta}), determining a different value of the third integrals in Eq. \ref{tauappf} and Eq. \ref{tauappjp}, $I_{\rm F}({\epsilon})$ and $I_{\rm JP}({\epsilon})$.

The ratio of the two functions $I_{\rm F}({\epsilon})$ and $I_{\rm JP}({\epsilon})$ for different values of $E$ and $E_{\rm LIV}=10^{19}$ GeV is in Fig. \ref{fig:appendix}. Clearly, for energies above the onset of LIV effects $I_{\rm JP}({\epsilon})>I_{\rm F}({\epsilon})$ for any $\epsilon$ (except for a small range around $\epsilon \sim \epsilon_{\rm th}$), resulting in $\tau_{\rm JP}(E)<\tau_{\rm F}(E)$.

\end{document}